\documentclass[iop]{emulateapj}

\newcommand{\HII}{{\ion{H}{2}}}

\newcommand{\OII}{[{\ion{O}{2}}]}

\newcommand{\OIIIHb}{[{\ion{O}{3}}]/H$\beta$}
\def\ratioR23{([\ion{O}{2}]~$\lambda$3727 +[\ion{O}{3}]~$\lambda\lambda$4959,5007)/H$\beta$}
\def\R23{${\rm R}_{23}$}

\newcommand{\Msun}{${\rm M}_{\odot}$}

\newcommand{\NII}{[{\ion{N}{2}}]}

\newcommand{\NIIHa}{[\ion{N}{2}]/H$\alpha$}
\newcommand{\SIIHa}{[\ion{S}{2}]/H$\alpha$}
\newcommand{\OIHa}{[\ion{O}{1}]/H$\alpha$}

\newcommand{\Hb}{{H$\beta$}}
\def\O4363{[{\ion{O}{3}}]~$\lambda$4363}
\newcommand{\OIII}{[{\ion{O}{3}}]}

\newcommand{\Ha}{{H$\alpha$}}

\def\L60{L$_{60}$}

\shorttitle{}
\shortauthors{}

\begin{document}

\title{The Cosmic BPT Diagram: Confronting Theory with Observations}

\author{Lisa J. Kewley}
\affil{Australian National University}
\affil {University of Hawaii}
\email {kewley@mso.anu.edu.au}

\author{Christian Maier}
\affil{University of Vienna}

\author{Kiyoto Yabe} 
\affil{Division of Optical and Infrared Astronomy, National Astronomical Observatory of Japan}

\author{Kouji Ohta}
\affil{Department of Astronomy, Kyoto University}

\author{Masayuki Akiyama}
\affil{Astronomical Institute, Tohoku University}

\author{ Michael A. Dopita}
\affil{Australian National University}
\affil{King Abdulaziz University}

\author{Tiantian Yuan}
\affil{Australian National University}


\begin{abstract}

We compare a large sample of galaxies between $0.5<z<2.6$ with theoretical predictions for how the optical diagnostic line ratios in galaxy ensembles change as a function of cosmic time.  We show that star forming galaxies at high redshift ($z>1.5$) are consistent with a model in which the ISM conditions are more extreme at 
high redshift than seen in the global spectra of local galaxies.  We speculate that global spectra of our high redshift galaxies may be dominated by \HII\ regions similar to the extreme clumpy, dense star-forming complexes in the Antennae and M82.    The transition to local-type conditions occurs between $0.8<z<1.5$.  We conclude that classification schemes developed for local samples should not be applied at high redshift ($z \geq 1.5$).  We use our theoretical models to derive a new redshift-dependent classification line that utilizes the standard optical diagnostic line ratios \OIIIHb\ and \NIIHa.  Our new line can be used to separate star-forming galaxies from AGN between $z=0$ to $z\sim 3.5$.   We anticipate that our redshift-dependent optical classification line will be useful for future large surveys with near-infrared multi-object spectrographs.  We apply our classification line to a sample of gravitationally lensed galaxies at $z\sim 2.5$.  Although limited by small numbers, we show that our classification line is consistent with the position of AGN that have been independently confirmed via other methods.
\end{abstract}


\section{Introduction}

Understanding how the fundamental physical properties of ensembles of galaxies change across cosmic time is one of the primary probes of galaxy evolution.
Spectral classification is crucial for this work.  Classification of galaxies according to their dominant power source is required for measuring the star-formation history of galaxies \citep[e.g.,][and many others]{Madau96,Lilly96,AHopkins06,Sobral13}, studying the main-sequence for star-forming galaxies \citep{Noeske07}, understanding the metallicity history of galaxies \citep{Kobulnicky04,Zahid11,Yuan13}, and disentangling the cosmic evolution of the starburst versus AGN fraction \citep{Almeida13}.

\citet{Baldwin81} first proposed the \NIIHa\ versus \OIIIHb\ diagnostic diagram (now known as the BPT diagram) to separate normal \HII\ regions, planetary nebulae, and objects photo-ionized by a harder radiation field.  A hard radiation field can be produced by either a power-law continuum from an AGN or from shock excitation.  \citet{Veilleux87} extended and refined this classification scheme.  They used theoretical photoionization models to inform the shape of the classification line between star-forming and AGN galaxies, and they added two new diagnostic diagrams that exploit the \SIIHa\ and \OIHa\ line ratios.  \citet{Kewley01a} combined stellar population synthesis and photoionization models to build the first purely theoretical classification scheme for separating pure AGN from galaxies containing star-formation.  

The Sloan Digital Sky Survey (SDSS) revolutionized optical classification schemes, allowing the shape and position of the local star-forming abundance sequence and the local starburst-AGN mixing sequence to be cleanly characterized \citep{Kauffmann03b, Kewley06a}.    

Spectral classification beyond $z>0.8$ has been difficult in the past.  The \NIIHa\ ratio is redshifted into the near-infrared (NIR) at $z\sim 0.5$, while at redshifts $z\sim 1.5$, all of the standard optical diagnostic ratios are in the NIR.  Fortunately, NIR multi-object spectrographs on 8-10m telescopes now allow statistically significant numbers of galaxies to be classified according to their dominant power-source.    

The \NIIHa\ and \OIIIHb\ ratios have now been observed for small numbers of galaxies at high redshift \citep[e.g.,][and references therein]{Erb06,Hainline09,Bian10,Rigby11,Yabe12}.   The majority of these galaxies exhibit larger \NIIHa\ and \OIIIHb\ ratios than seen in local star-forming galaxies.  \citet{Brinchmann08b} used stellar population synthesis and photoionization models to show that local galaxies with large \NIIHa\ and \OIIIHb\ ratios have a larger ionization parameter than the general star forming population.   They suggest that higher electron densities and a larger escape fraction of H-ionizing photons may be responsible for the larger ionization parameter.

In \citet{Kewley13a}, we combined the chemical evolution predictions of cosmological hydrodynamic simulations with stellar evolutionary synthesis and photoionization models to predict how the optical emission-line ratios in star-forming and AGN populations will change with redshift for four different model scenarios.  We showed that the position of star-forming galaxies may change significantly as a function of redshift if the ISM conditions and/or the stellar ionizing (EUV) radiation field within the star-forming population changes with cosmic time. 
 
In this letter, we compare our new theoretical predictions with a statistically significant sample of galaxies between $0.8<z<3$ for which near-infrared spectra are available.  We speculate on the star-forming conditions in high redshift galaxies, and we develop a new optical classification scheme that can be applied between $0<z<3.5$.  We adopt the flat $\Lambda$-dominated cosmology from the 7 year WMAP experiment \citep[$h=0.72$, $\Omega_{m}=0.29$;][]{Komatsu11}.

\section{Sample }\label{Sample}

Our sample consists of 76 galaxies from magnitude-limited and lensed samples with $z\gtrsim 0.5$ and measurable \NII, \Ha, \OIII, and \Hb\ emission-lines ($S/N \geq 3\sigma$).  These samples contain the most massive ($M>10^{9.5}$\Msun) actively star-forming galaxies at their respective redshifts and may be missing low mass, low metallicity galaxies, characterized by low \NIIHa\ and high \OIIIHb\ ratios.   This selection is sufficient to test the applicability of the local classification scheme at intermediate and high redshift because the most massive, actively star-forming galaxies lie closest to the starburst-AGN mixing sequence of the BPT diagram \citep[e.g.,][]{Yuan10}.   Extending our results to the general population of high redshift star-forming galaxies (i.e. fully characterizing the star-forming abundance sequence with redshift) requires galaxies to be sampled over $\sim 4$ magnitudes of stellar mass and is not possible with current samples.

At $z\sim 0.8$, we use galaxies from the zCOSMOS spectroscopic survey \citep{Lilly07,Lilly09} of the $1.5 {\rm deg}^{2}$ COSMOS field \citep{Scoville07a}. The zCOSMOS-bright sample contains VIMOS spectra of $\sim 20 000$ galaxies with $I_{AB} \leq 22.5$ between $0<z<1.4$, yielding \OII, \Hb\ and \OIII\ line fluxes for galaxies between $0.5<z<0.92$. Near infrared VLT-ISAAC spectroscopy allowed the additional measurements of \Ha\ and \NII\ for a sample of 18 galaxies between $0.5<z<0.92$ and 2 galaxies at $z\sim 2.5$ (Maier et al. 2013, in prep).  For our analysis of the \OIIIHb\ line ratio, we also use the 22 zCOSMOS-bright galaxies with \OIIIHb\ measurements but $3\sigma$ upper limits on \NII.  This supplementary sample with \NII\ upper limits allows us to probe lower metallicity, lower stellar mass ($10^{9.5}-10^{10.5}$\Msun) galaxies for our \OIIIHb\ ratio analysis.  The zCOSMOS sample excludes AGN identified with X-ray data.
We supplement the zCOSMOS data with 6 galaxies at $z\sim 0.8$  from the DEEP2 Galaxy Redshift Survey \citep{Davis03,Faber07} observed by \citet{Shapley05} and \citet{Liu08}.  We also include 1 gravitationally lensed galaxy observed by \citet{Christensen12}.  

At $z\sim 1.5$, we use 87 galaxies observed by \citet{Yabe12} and Yabe et al. (2013, MNRAS, submitted) from the Subaru-XMM Deep Survey (SXDS) \citep{Furusawa08} and the UKIDSS Ultra Deep Survey (UDS) \citep{Lawrence07}.   The SXDS has limiting magnitudes of $\sim 27$ in the $B,V,R_{C},i'$ bands and a limiting magnitude of $26$ in the $z'$ band, while the UDS is more shallow (limiting magnitudes of $\sim 24-25$ in the $J,H,K$ bands).  Yabe et al. selected galaxies for near-infrared observations to satisfy: $K_{s} < 23.9$~mag, stellar mass $M_{*} \geq 10^{9.5}$~\Msun, and expected \Ha\ flux $F_{H\alpha} > 5.0 \times 10^{-17}$~erg~s$^{-1}$~cm$^{-2}$.   The sample excludes X-ray sources, thus avoiding luminous AGN.  Of the 87 Yabe et al. galaxies, 20 objects have 3$\sigma$ detections for all 4 BPT emission-lines.  The remaining galaxies have upper limits on either \OIII, \NII, or \Hb, which we use for our \OIIIHb\ line analysis.  We supplement the Yabe et al. data with 6 DEEP2 galaxies between $1.36\leq z\leq 1.50$ observed by   \citet{Shapley05} and \citet{Liu08}, and 6 gravitationally lensed galaxies by \citet{Rigby11}, \citet{Yuan13}, and \citet{Christensen12}.   

At $z\sim 2.5$, we use 5 galaxies from the SINS sample \citep{Forster09}, 2 zCOSMOS galaxies (Maier et al. 2013, in prep), the low metallicity $L_{*}$ galaxy (Q2343-BX418) from \citet{Erb10}, and 16 gravitationally lensed galaxies from the compilations of \citet{Richard11}, \citet{Yuan13}, and \citet{Jones13}.  Of these, a total of 19 galaxies have $>3\sigma$ detections of the \NII, \Ha, \OIII, \Hb\ emission-lines.  Gravitational lensing boosts the luminosity of galaxies by $10-30 \times$, allowing lower luminosity, lower metallicity galaxies to be sampled compared with magnitude-limited surveys.  The magnification (and hence the limiting magnitude) differs for each lensing source, depending on the geometry of the lens and the background galaxy.  Although the lensing magnification factor does not affect the emission-line ratios, the distortion of the galaxy image causes aperture effects.  \citet{Hainline09} show that this effect is likely to be small ($0.05 - 0.2$~dex in the \OIIIHb\ and \NIIHa\ line ratios).

The combination of magnitude-limited and lensed galaxies gives a total of  25, 32, and 19 galaxies at $z=0.8,1.5,2.5$ with measured \NIIHa\ and \OIIIHb\ line ratios.  We also analyse a further 97 galaxies from these samples with upper limits on either \OIII, \NII, or \Hb.   

\section{Testing the Cosmic BPT Diagram }\label{Cosmic_BPT}

In \citet{Kewley13a}, we combine the chemical evolution predictions from \citet{Dave11b} with the Starburst99 \& Pegase2 evolutionary synthesis models \citep{Leitherer99,Fioc99} and our Mappings IV photoionization code \citep{Dopita13}.  Briefly, we use the Dave et al. chemical evolution predictions for the star-forming gas in $M_* >10^{9}$~\Msun\ galaxies with momentum-conserving winds ({\it vzw} in Dave et al).   We use the instantaneous and continuous burst models from Starburst99 and Pegase2 with a Salpeter Initial Mass Function, and the Pauldrach/Hillier stellar atmosphere models.  The resulting stellar population spectra are used as the ionizing source in our Mappings IV photoionization code, with a spherical nebular geometry.   Mappings IV uses a Kappa temperature distribution which is more suitable for a turbulent ISM than a Stefan-Boltzmann distribution \citep{Nicholls12}.  For AGN, we use the dusty radiation-pressure dominated models of \citet{Groves04a}.  Further details of all model used are given in \citet{Kewley13a} and \citet{Dopita13}.   Our combination of these models produces theoretical predictions for how the optical emission-line ratios will appear for galaxy samples at different redshifts, based on two limiting assumptions for star-forming galaxies and AGN:

\begin{itemize}

\item Star forming galaxies at high redshift ($z=3$) may have ISM conditions and/or an ionizing radiation field that are either the same as local galaxies ({\it normal ISM conditions}) or else are more extreme than local galaxies ({\it extreme ISM conditions}).  Extreme conditions in star-forming galaxies can be produced by a larger ionization parameter (by a factor of 2) and a denser interstellar medium (by a factor of 10), and/or an ionizing radiation field SED that contains a larger fraction of photons able to ionize O$^{+}$ into O$^{++}$ (i.e. energy $>35.12$~eV) relative to the number of H-ionizing photons.

\item The AGN narrow line region at high redshift may either have already reached the level of enrichment observed in local galaxies ({\it metal-rich}), or else it has the same metallicity as the surrounding star-forming gas.  In the latter case, the AGN narrow-line region at high redshift would be more {\it metal-poor} than local AGN narrow-line regions.  The metal-rich AGN case is equivalent to high-$z$ AGN galaxies containing steeper metallicity gradients than seen in local galaxies, while the metal-poor AGN case is equivalent to high-$z$ AGN galaxies containing flatter metallicity gradients than seen in local galaxies.  

\end{itemize}

These limiting assumptions are described in detail in \citet{Kewley13a}.  In this work, we focus on the position of the star-forming abundance sequence rather than the position of the AGN mixing sequence.  

\begin{figure}[!h]
\epsscale{1.0}
\plotone{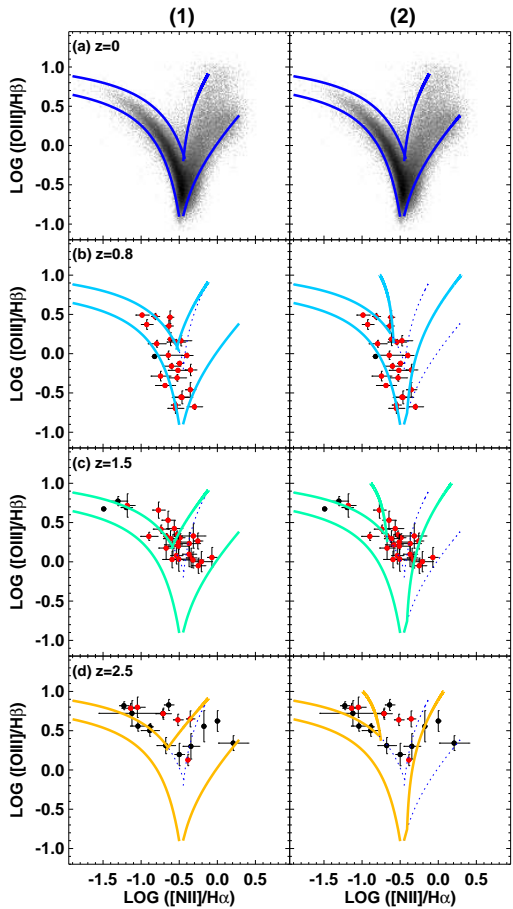}
\caption[Cosmic_BPT_nolims]{The Cosmic BPT Diagram for four different redshift ranges (rows) and our model scenarios (1) and (2). In each panel, solid lines show our theoretical predictions for the position of the star-forming abundance sequence (left curves) and the AGN mixing sequence (right curves).  Black and red circles indicate the positions of gravitationally lensed and magnitude-limited survey galaxies, respectively.  Blue dotted lines indicate the boundaries at $z=0$, for comparison.
\label{Cosmic_BPT_nolims}}
\end{figure}

\begin{figure}[!h]
\epsscale{1.0}
\plotone{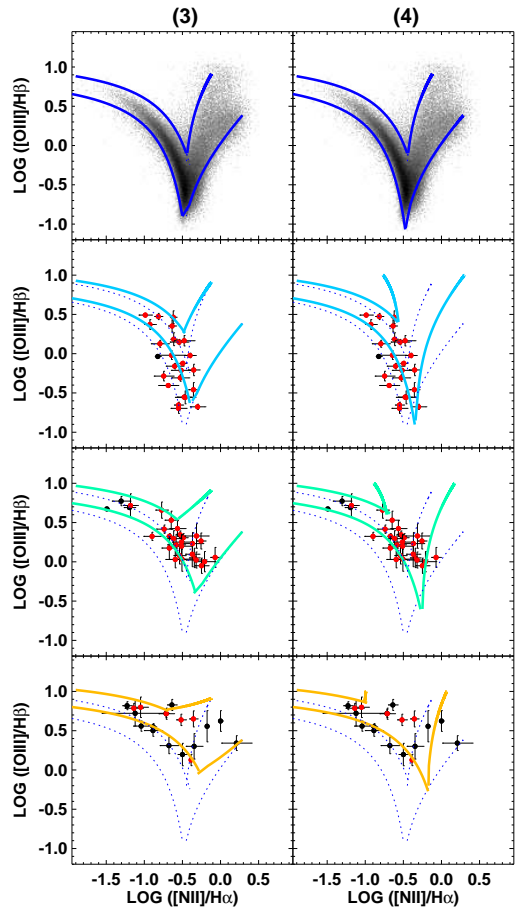}
\caption[Cosmic_BPT_nolims_2]{As in Figure~1, but for our model scenarios (3) and (4). Data at $z>1$ are most consistent with scenarios (3) and (4), in which star-forming galaxies have more extreme ISM conditions at high redshift.  
\label{Cosmic_BPT_nolims_2}}
\end{figure}

In Figures~\ref{Cosmic_BPT_nolims} and \ref{Cosmic_BPT_nolims_2}, we compare our samples with our model predictions.  The columns indicate the four different model scenarios for the star-forming (SF) galaxies and the AGN narrow-line region (NLR) that are obtained by our two sets of limiting assumptions: 

\begin{enumerate}
\item Normal SF/ISM conditions; metal-rich AGN NLR at high-z (Figure 1, left panel).
\item Normal SF/ISM conditions; metal-poor AGN NLR at high-z (Figure 1, right panel).
\item Extreme SF/ISM conditions at high-z; metal-rich AGN NLR at high-z (Figure 2, left panel).
\item Extreme SF/ISM conditions at high-z; metal-poor AGN NLR at high-z (Figure 2, right panel).
\end{enumerate}

In Figures~\ref{Cosmic_BPT_nolims} and \ref{Cosmic_BPT_nolims_2}, galaxies at $z\sim 0.8$ span a similar range of \OIIIHb\ ratios to local star-forming galaxies.  However, at $z>1$, all galaxies in our samples (both lensed and magnitude-limited) are offset towards larger \OIIIHb\ ratios than local star-forming galaxies over the same \NIIHa\ range (i.e the same metallicity range).   This result is consistent with previous work on smaller samples \citep[e.g.,][]{Shapley05,Lehnert09,Rigby11,Yabe12}.  

\begin{figure}[!h]
\epsscale{1.2}
\plotone{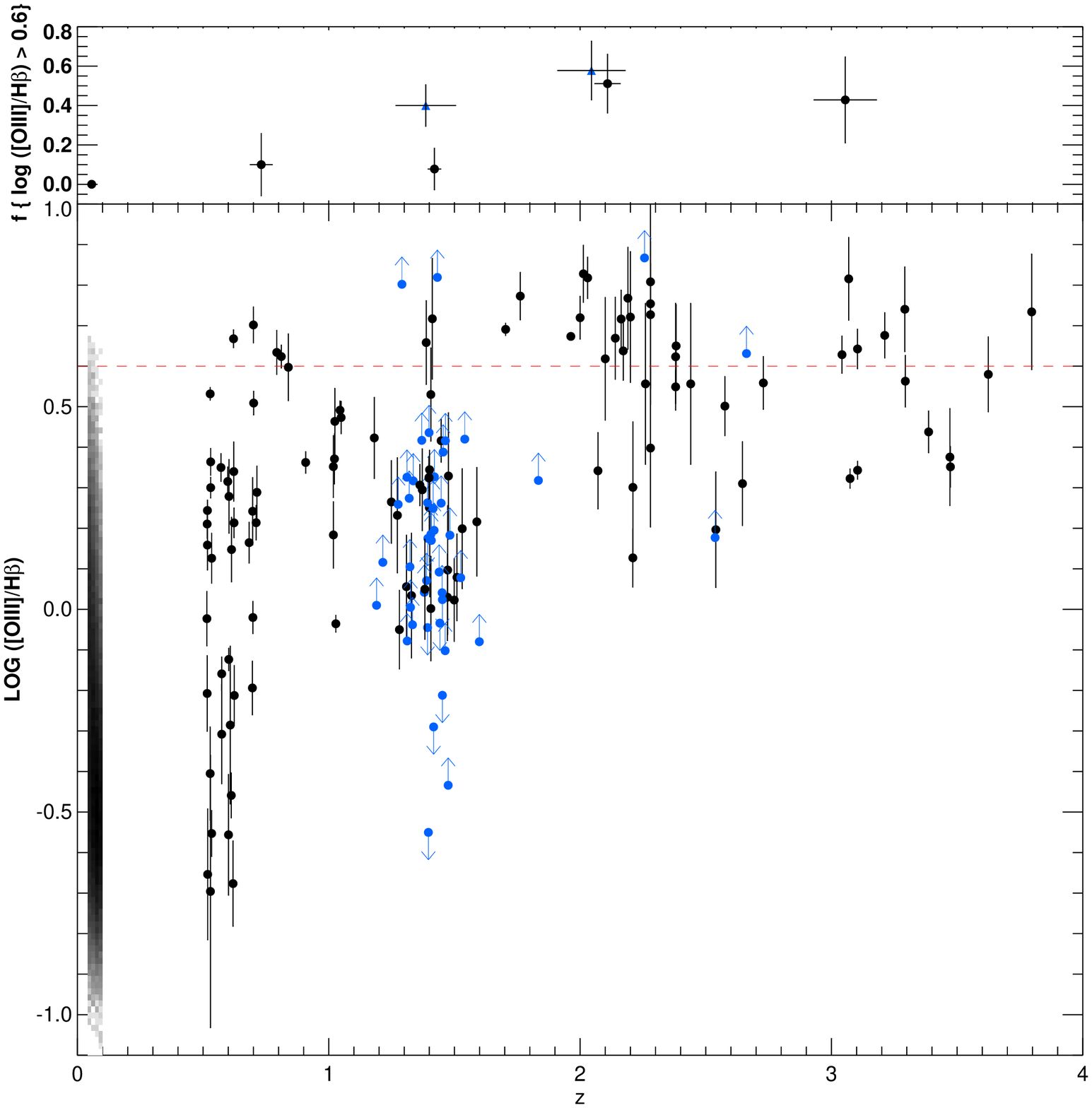}
\caption[OIIIHb_z]{ Lower panel: The \OIIIHb\ ratio versus redshift, including \OIII\ and \Hb\ upper limits.  Top panel: Fraction of galaxies with $\log$(\OIIIHb)$>0.6$ for 5 redshift intervals.  The fraction that would be obtained if all galaxies with \OIIIHb\ lower limits $\log$(\OIIIHb)$<0.6$ truly have $\log$(\OIIIHb)$<0.6$ (black circles) or if they have $\log$(\OIIIHb)$>0.6$ (blue triangles) are shown.  The rise in $\log$(\OIIIHb)$>0.6$ with time from $z\sim0$ is statistically significant for $z\gtrsim 2$.
\label{OIIIHb_z}}
\end{figure}

We investigate whether this offset is due to observational detection limits in Figure~\ref{OIIIHb_z} (lower panel), which gives the \OIIIHb\ ratio as a function of redshift.  One advantage of this diagram is that we can include galaxies without measurable \NII\ and \Ha, allowing us to include higher redshift samples.  We include the 4 Lyman Break galaxies from \citet{Pettini01}, 7 Lyman Break galaxies from \citet{Maiolino08}, and 3 lensed galaxies from \citet{Richard11} with measured \OIIIHb\ ratios (to the 3$\sigma$ level).  Figure~\ref{OIIIHb_z} also shows galaxies with $3\sigma$ upper limits on \OIII\ or \Hb, where available.  The \OIIIHb\ ratios are systematically larger for galaxies at $z>1$, even when upper limits are taken into account.   The majority of galaxies (34/47; 87\%) with \OIIIHb\ limits have \OIII\ detections but no \Hb\ detections (i.e., {\it lower limits} on the \OIIIHb\ ratio), strengthening the rise of \OIIIHb\ with redshift.   

The \OIIIHb\ upper envelope also rises with redshift such that galaxies have larger \OIIIHb\ ratios on average at $z>1$ than galaxies at lower redshifts, over the same \NIIHa\ (or metallicity) range.  The fraction of galaxies with \OIIIHb$> 0.6$ rises with redshift (Figure~\ref{OIIIHb_z} top panel).  This rise is not caused by detection bias because we are able to detect \OIIIHb\ for AGN with \OIIIHb$> 0.6$ in the same observational samples at $z<1$.

We propose that the rise in \OIIIHb\ with redshift is caused by either (or a combination of): (1) a larger ionization parameter and a denser interstellar medium at high redshift, and/or an ionizing SED that contains a larger fraction of O$^{+}$-ionizing to H-ionizing photons.   The larger ionization parameter may be related to the large specific star formation rate observed at high redshift \citep[e.g.,][]{Noeske07}.   We note that in the local galaxies M82 and the Antennae, extremely high ionized gas densities and ionization parameters are found in clumpy, dense star forming complexes  \citep{Smith06,Snijders07}.  The densities and ionization parameters measured in these dense complexes are similar to those observed at high redshift \citep[e.g.][and references therein]{Rigby11}.   \citet{Snijders07} shows that a geometrical model in which several individual star-forming clumps are embedded in a giant molecular cloud can reproduce these extreme ISM conditions.  
We speculate that global spectra of our high redshift galaxies may be dominated by \HII\ regions similar to the clumpy, dense star-forming regions in the Antennae and M82.  It is also possible that the \HII\ regions at high redshift are predominantly matter bounded.  In this case, the effective ionization parameter from our radiation bounded models would appear larger \citep[see][for a discussion]{Kewley13a}.     In a follow-up paper (Kewley et al. in prep), we investigate these scenarios in detail.

\section{New Redshift-Dependent Classification Scheme} \label{Classification}

Figure~\ref{Cosmic_BPT_nolims_2} indicates that the current optical spectral classification schemes are not suitable for classifying galaxies at $z>1$.  Star-forming galaxies are best-fit by the theoretical models described in Scenarios 3 and 4.  In both of these scenarios, the star-forming abundance sequence can be reproduced by either (1) an ionizing radiation field with a larger fraction of O$^{+}$-ionizing to H-ionizing photons, and/or (2) a combination of a larger ionization parameter and a high electron density.   We use the upper curve of our theoretical abundance sequence to define semi-empirically how the position of the classification line may change with redshift:

\begin{eqnarray}
\log([OIII]/H\beta) & =  & \frac{0.61}{( \log([NII]/H\alpha) -0.02 - 0.1833 * z )}  \nonumber \\ 
                            &    &  +  1.2 + 0.03 * z  \label{Eq_class}
\end{eqnarray}

where \OIIIHb\ uses the \OIII~$\lambda 5007$ line and \NIIHa\ uses the \NII~$\lambda 6584$ line.  
Equation~\ref{Eq_class} can be used to separate purely star-forming galaxies from galaxies containing AGN, and is based on the following assumptions: 

\begin{itemize}
\item The shape of the ionizing radiation field and the ISM conditions in star-forming galaxies evolves only marginally ($<0.1$ dex) with redshift until $z\sim1$, consistent with our findings in Figure~\ref{Cosmic_BPT_nolims_2}.  This assumption may only hold for galaxies in the mass range of our samples (${\rm M}_{*} > 10^9 {\rm M}_{\odot}$).
The star-forming conditions in galaxies with lower stellar masses may still be evolving between $z=1$ and $z=0$. 
If star-forming conditions are more extreme in low stellar mass galaxies in this redshift range, then these galaxies will lie above and to the right of the line given by equation~\ref{Eq_class}.  In this case, our optical spectral classification line would need to include a stellar mass term.  

\item The shape of the ionizing SED and/or the ISM conditions in star-forming galaxies evolves towards more extreme conditions between $1<z<3$.  Figure~\ref{Cosmic_BPT_nolims_2} suggests that such a change does occur, at least in the most massive galaxies (${\rm M}_{*} > 10^9$~\Msun).  In Figures~\ref{Cosmic_BPT_nolims} and \ref{Cosmic_BPT_nolims_2}, we have binned galaxies into two redshift ranges with median redshifts of $z=1.5$ and $z=2.5$ to provide a sufficient number of galaxies in each redshift interval for statistically significant conclusions.  This small number of redshift intervals allows us to coarsely fit the evolution of the abundance sequence between $0.8<z <2.5$.  A substantially larger sample ($\gtrsim 200$ galaxies) would allow one to divide the sample into a larger number of small redshift intervals, allowing for a more robust fit in this redshift range. 
\end{itemize}

We note that our chemical evolution assumption affects the position of the intersection between the star-forming abundance sequence and the AGN mixing sequence, but does not affect the shape or location of the classification line.

\begin{figure}[!h]
\epsscale{1.0}
\plotone{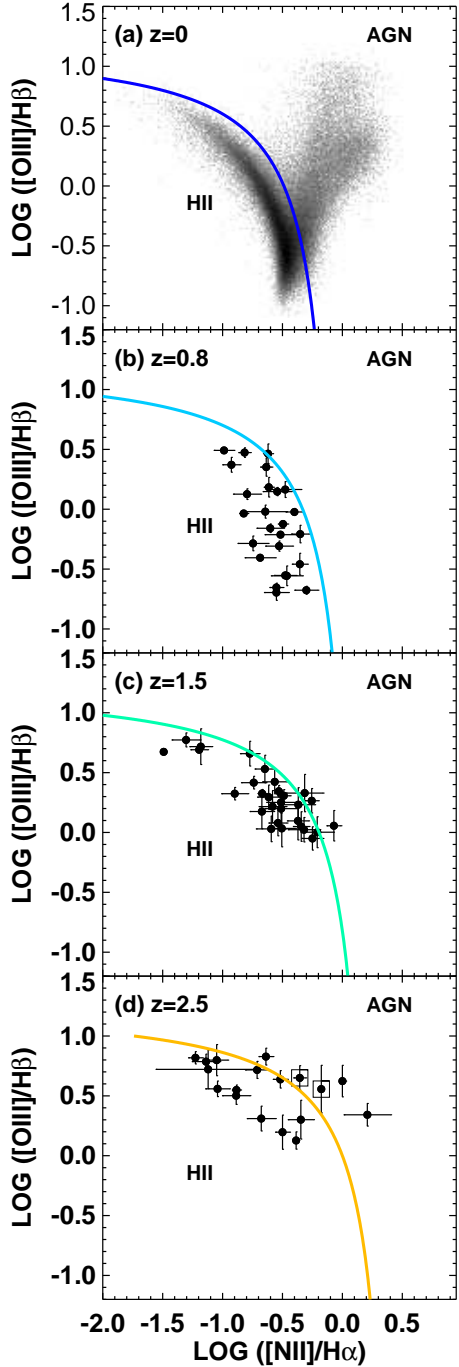}
\caption[z_class]{Our new theoretical redshift-dependent classification scheme for four redshifts.  Galaxies at z$\sim 2.5$ with additional evidence for AGN are shown with a square outline.
\label{z_class}}
\end{figure}

In Figure~\ref{z_class} we show the classification line from equation~\ref{Eq_class} for our 4 redshift ranges.  The zCOSMOS-bright ($z\sim 0.8$) and Yabe ($z\sim 1.5$) NIR samples excluded bright AGN using X-ray data.  The location of the star-forming galaxies in these samples is consistent with our new classification line, within the errors ($\pm 0.1$~dex).  

Of the 5 $z\sim 2.5$ galaxies that we classify optically as AGN, two galaxies (MACS J09012240+1814321 and BzK 15504; square symbols in Figure~\ref{z_class}) show additional evidence for an AGN.   UV spectroscopy led \citet{Diehl09} to suggest that MACS J0901 contains an AGN.  Although dominant in the UV, the AGN does not make an energetically significant contribution to the global mid-infrared spectrum \citep{Fadely10}.  \citet{Lehnert09} suggest that BzK15504 contains an AGN based on its strong \OIHa\ emission-line ratio.  The remaining 3 optically classified AGN are A1835 \citep{Richard11}, MACS J1148 \citep{Jones13}, and Q2343-BX389 \citep{Lehnert09}.  These 3 galaxies do not have sufficient ancillary data to confirm or rule out an AGN at other wavelengths.   Follow-up of these 3 galaxies at X-ray, UV, and/or mid-infrared wavelengths will facilitate testing of our new classification scheme.

\section{Conclusions}

We have presented the first comparison between theoretical predictions of the cosmic evolution of the BPT diagram and near-infrared spectroscopic observations of active galaxies between $0.8<z<2.5$.  We show that star-forming galaxies at high redshift are consistent with a model in which the ISM conditions are more extreme at high redshift.  These extreme conditions may manifest in either (or a combination of) a larger ionization parameter, a larger electron density, and/or an ionizing radiation field with a larger fraction of O$^{+}$-ionizing to H-ionizing photons.  We speculate that the global spectra of high redshift galaxies may be dominated by complexes of 
star-forming clusters embedded within giant molecular clouds, as seen in the most extreme star forming regions in the Antennae and M82.

Due to the change in ISM conditions in star-forming galaxies with redshift, current optical classification methods based on local samples are not reliable beyond $z>1$. 
We present a new redshift-dependent optical classification line for the \OIIIHb\ versus \NIIHa\ diagnostic diagram that accounts for how the structure of the ISM changes with redshift.  We show that the position of independently confirmed AGN is consistent with our classification line at $z\sim 2.5$.  

In an upcoming paper, we will further investigate the cause of the extreme ISM conditions in high redshift star-forming galaxies.

\acknowledgments
L.K. gratefully acknowledges the referee for useful comments, the support of an ARC Future Fellowship, ARC Discovery Project DP130103925, and the ANU CHELT Academic Women's Writing Workshop.  This research used NASA's Astrophysics Data System Bibliographic Services and the NASA/IPAC Extragalactic Database (NED).


\end{document}